\documentclass[10pt,conference,a4paper]{IEEEtran}
\usepackage[utf8]{inputenc}

\usepackage{blindtext, graphicx}
\usepackage[utf8]{inputenc}
\usepackage[english]{babel}
\usepackage{csquotes}
\usepackage{wrapfig}
\usepackage{float}
\usepackage{verbatim}
\usepackage{subcaption}
\usepackage{import}
\usepackage{dirtytalk}
\usepackage{subfiles}
\usepackage{array}
\usepackage{multirow}

\usepackage{lscape}
\usepackage{algorithm}
\usepackage{algorithmic}
\usepackage{rotating}
\usepackage{listings}
\usepackage{booktabs}
\usepackage[table,xcdraw]{xcolor}

\usepackage{times}

\DeclareGraphicsExtensions{.png,.eps,.ps,.pdf}

\usepackage{url}
\usepackage{hyperref}

\usepackage[most]{tcolorbox}

\usepackage{multirow}
\usepackage{tabularx}
\usepackage{makecell}
\usepackage{svg}

\usepackage{pifont}
\definecolor{ao}{rgb}{0.0, 0.5, 0.0}
\definecolor{amber}{rgb}{1.0, 0.49, 0.0}

\let\labelindent\relax
\usepackage{enumitem}

\usepackage{lscape} 

\graphicspath{{images/}}

\usepackage{tikz} % For \foreach used for Orcid icon
\newcommand{\orcidicon}{\includegraphics[width=0.32cm]{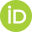}}
\foreach \x in {A, ..., Z}{%
\expandafter\xdef\csname orcid\x\endcsname{\noexpand\href{https://orcid.org/\csname orcidauthor\x\endcsname}{\noexpand\orcidicon}}
}

\begin{document}

\title{Anticipating Adversary Behavior in DevSecOps Scenarios through Large Language Models}

\author{\IEEEauthorblockN{
\orcidA{}Mario Mar\'in-Caballero$^1$,
\orcidB{}Miguel Betancourt Alonso$^2$,
\orcidC{}Daniel D\'iaz-L\'opez$^{1,2}$\\
\orcidD{}Angel Luis Perales G\'omez$^{3}$
\orcidF{}Pantaleone Nespoli$^{1}$
\orcidE{}Gregorio Mart\'inez P\'erez$^{1}$
}
\IEEEauthorblockA{$^1$Department of Information and Communications Engineering, University of Murcia, 30100, Murcia, Spain\\
\{mario.m.c, danielorlando.diaz, pantaleone.nespoli, gregorio\}@um.es}
\IEEEauthorblockA{$^2$School of Engineering, Science and Technology, Universidad del Rosario, Bogot\'a, Colombia\\
\{miguels.betancourt, danielo.diaz\}@urosario.edu.co}
\IEEEauthorblockA{$^3$Department of Computers Engineering and Technology, University of Murcia, 30100, Murcia, Spain\\
angelluis.perales@um.es}

}


\maketitle

\begin{abstract} %Miguel
%Importance of protect DevSecOps scenarios as part of a strategy of cyber defense %Main contribution
The most valuable asset of any cloud-based organization is data, which is increasingly exposed to sophisticated cyberattacks. Until recently, the implementation of security measures in DevOps environments was often considered optional by many government entities and critical national services operating in the cloud. This includes systems managing sensitive information, such as electoral processes or military operations, which have historically been valuable targets for cybercriminals. Resistance to security implementation is often driven by concerns over losing agility in software development, increasing the risk of accumulated vulnerabilities. Nowadays, patching software is no longer enough; adopting a proactive cyber defense strategy, supported by Artificial Intelligence (AI), is crucial to anticipating and mitigating threats. Thus, this work proposes integrating the Security Chaos Engineering (SCE) methodology with a new LLM-based flow to automate the creation of attack-defense trees that represent adversary behavior and facilitate the construction of SCE experiments based on these graphical models, enabling teams to stay one step ahead of attackers and implement previously unconsidered defenses. Further detailed information about the experiment performed, along with the steps to replicate it, can be found in the following repository: \url{https://github.com/mariomc14/devsecops-adversary-llm.git}.

\end{abstract}


{\bf Contribution type:}  {\it  Original research }

\section{Introduction}\label{intro} 

%Paragraph' subject: Analysis of Adversary Behavior as part of a cyber defense strategy
Cyber defense refers to the set of security measures and strategies against cyberattacks~\cite{srinivas2019government} and can be implemented either actively or passively. Historically, private and public corporations have mostly relied on passive measures, focusing on patching vulnerabilities. However, active cyber defense allows for a broader perspective and the consideration of a wide range of interaction scenarios between individuals, which has proven to be a more effective strategy~\cite{serban2023malware}. This shift from reactive to proactive approaches has led organizations to seek methodologies that can systematically strengthen their security posture.

One such methodology that supports the implementation of a proactive cyber defense strategy is Security Chaos Engineering (SCE), which improves security and tests the resilience of cloud-deployed architectures through fault injection and controlled experimentation~\cite{9133399}. SCE represents a paradigm shift in how organizations approach security, moving from merely responding to threats toward actively testing defenses before attackers can exploit them. This proactive approach aligns perfectly with modern software development practices that emphasize agility and continuous improvement.

DevOps, which refers to the agile and collaborative integration of development and operations teams ~\cite{7458761}, embodies these modern practices. When security tasks are integrated from the early stages of development through to software delivery, the practice evolves into DevSecOps. This integration creates a comprehensive environment where security becomes an inherent part of the development lifecycle rather than an afterthought or separate concern, thereby strengthening an organization's overall defensive posture.

An optimal cyber defense strategy in a DevSecOps scenario should integrate vulnerability identification, prioritize threats based on their likelihood of occurrence, explore various attack scenarios while anticipating adversary behavior, and implement automated and effective countermeasures. This holistic approach requires sophisticated tools and methodologies that can model complex attacker-defender dynamics within fast-paced development environments.

%Paragraph' subject: Anticipate adversary behavior by traditional methods
Attack-defense trees represent one traditional approach to addressing this need, offering a comprehensive visual framework for cybersecurity analysis by combining offensive and defensive actions in a hierarchical structure. By extending traditional attack trees to include defensive countermeasures, these models enable a dynamic representation of the attacker-defender interplay. This structured approach facilitates improved risk assessment, interdisciplinary collaboration, and strategic decision-making in addressing complex, multi-stage security threats within DevSecOps pipelines.

However, when using these methods (considered traditional approaches for anticipating adversary behavior in cyberspace), their effectiveness is limited by the creativity and knowledge of the security teams implementing them. This limitation creates a significant challenge in maintaining both security and agility in DevSecOps environments, as human-centered analysis can become a bottleneck in rapid development cycles while still potentially missing novel attack vectors.

%Paragraph' subject: LLM's
In response to these limitations, the state-of-the-art in Large Language Models (LLMs) has emerged as a promising solution by enabling the automation of tasks requiring human language comprehension and generation. Some LLMs, such as RepairLlama~\cite{silva2023repairllama}, which corrects code errors, and Hackmentor~\cite{zhang2024when}, which applies fine-tuning to pre-existing LLM models, have been trained directly on cybersecurity data, making them specialized models in this field. Others, like ChatGPT~\cite{OpenAI2024}, are pre-trained on general data and later adapted using fine-tuning techniques or methods such as in-context learning and chain-of-thought reasoning to tackle complex cybersecurity tasks~\cite{Yang2024}.

Unlike RepairLlama and Hackmentor, the great potential of ChatGPT lies not only in its ability to correct code errors, but also in its outstanding performance across various cybersecurity tasks, such as autonomous vulnerability exploration and the discovery of new threats~\cite{Bedoya2024}, without requiring a prior transfer learning process, making it readily available for immediate use.

%Conclusion of the intro
The strategic implementation of LLMs to forecast adversarial behavior patterns (specifically prospective attack procedures, their sequential execution patterns (e.g., branch sequencing within attack-defense tree frameworks), and associated cost-benefit considerations) holds significant potential to mitigate cognitive biases inherent in conventional security testing methodologies within DevSecOps pipelines. Furthermore, such an approach facilitates the systematic exploration of novel attack vectors while refining vulnerability detection and mitigation strategies. Operational integration of such predictive systems within cyber command frameworks could enhance proactive threat modeling, enabling stakeholders to preemptively identify and neutralize emerging threats targeting cloud-based national critical infrastructure assets.

%Contributions
Thus, the main contributions of this paper are summarized as follows:
\begin{itemize}
    \item A new LLM-based flow for generating attack-defense trees that represent adversary behavior.
    \item A comparative analysis of attack-defense tree generation is conducted using different general-purpose LLMs, providing the models with a set of prompts within a DevSecOps and cyber defense context to evaluate their ability to generate realistic attack structures with a clear hierarchy aligned with the proposed attack objective.
    \item The definition of key metrics for assessing the quality of attack-defense trees generated by different models. These metrics ensure that the trees are based on real-world attacks and countermeasures, maintain a clear hierarchical structure, and align with a well-defined attack objective.
    \item The validation of the proposal through the evaluation of some generated attack-defense trees, and the implementation of a SCE experiment that allow to test the feasibility of the generated paths from the perspective of a realistic adversary.
\end{itemize}

%Summary of sections
This paper is structured as follows: Section~\ref{sota} explores the LLMs used for cybersecurity and cyber defense purposes. Then, Section~\ref{method} describes our proposed method for generating and evaluating attack-defense trees using general-purpose LLM models. Next, in Section~\ref{experiments}, the best-obtained attack-defense trees are used to construct and execute our experiments. Finally, Section~\ref{conclusions} presents the conclusions and analyses potential future research directions to improve our proposal.

\section{State of the art}\label{sota} 

%Intro %Related work 1 (General use of LLMs)
This section describes recent advancements in the use of LLMs for cybersecurity and new frontiers in proactive security analysis. In this sense, the work proposed by Yadong Zhang \textit{et al.} \cite{zhang2024llm} provides a comprehensive survey on leveraging LLMs for strategic reasoning—a complex cognitive ability involving decision-making in multi-agent, dynamic, and uncertain environments. The study systematically explores LLMs' applications, methodologies, and evaluation in strategic scenarios, emphasizing their growing significance in multi-agent interactions and decision-making.

%Related work 2 (LLM to predict IoT intrusions)
Moreover, Alaeddine Diaf \textit{et al.} \cite{diaf2024beyond} present a novel framework for IoT networks that leverages LLMs and Long Short-Term Memory (LSTM) models to predict intrusions by analyzing network packets. The framework uses a fine-tuned Generative Pre-trained Transformer (GPT) model to predict network traffic and a fine-tuned Bidirectional Encoder Representations from Transformers (BERT) to evaluate the predicted traffic. An LSTM classifier then identifies malicious packets among these predictions.

%Related work 3 (LLMs to analyze TTPs employed by malware)
Another key research stream is the use of LLMs to discover attack procedures. To this extent, the work proposed by Ying Zhang \textit{et al.} \cite{zhang2024tactics} explores the development and deployment of GenTTP, a zero-shot framework that employs LLMs to analyze and automatically generate Tactics, Techniques, and Procedures (TTPs) for interpreted malware specific to Open-Source Software (OSS) ecosystems (e.g., PyPI, NPM). Unlike traditional malware analysis, which relies on manual reverse engineering or dynamic sand-boxing, GenTTP focuses on software supply chain (SCS) attacks, analyzing malicious OSS packages designed to infiltrate systems via deceptive metadata and stealthy code execution.

%Related work 4 (LLM to improve BAS functions)
Furthermore, Lingzhi Wang \textit{et al.} \cite{wang2024sandsmansionssimulatingattack} introduce Aurora, a semi-automated Breach and Attack Simulation (BAS) system designed to construct multi-step cyberattack chains efficiently. Aurora leverages LLMs, such as GPT, along with the Planning Domain Definition Language (PDDL), to address challenges in traditional BAS, such as limited action space, expertise requirements, and insufficient attack chain realism. Aurora automates the simulation of full cyberattack chains by analyzing attack tool documentation, threat intelligence reports, and real-world techniques.

%Related work 5 (LLM for revision/syntesis of cybersec content)
The goal of the current paper is to improve the defenses and resilience of systems focused on cyber defense environments. Under this paradigm, Christoforus Yoga Haryanto \textit{et al.} ~\cite{haryanto2024contextualized} demonstrate how contextualized AI can accelerate strategic decision-making by conducting an automated literature survey using LLMs. Contextualized AI significantly enhances cyber defense by accelerating threat intelligence synthesis, enabling strategic decision support, improving human-AI collaboration, and facilitating proactive defense posturing. 

%Related work 6 (LLM-based agents for cyberdefense purposes)
Also, Johannes F. Loevenich \textit{et al.} \cite{10773923} focus on the design and training of robust Autonomous Cyber Defense (ACD) agents for use in military networks. It introduces an architecture that combines a hybrid AI model incorporating Multi-Agent Reinforcement Learning (MARL), LLMs, and a rule-based system. These components are organized into blue and red agent teams distributed across network devices. The primary objective is to automate essential cybersecurity functions, including monitoring, detection, and mitigation, thereby enhancing the ability of cybersecurity professionals to safeguard critical military infrastructure.  

%Conclusion of SoA
The previous related works highlight LLMs' transformative potential in cybersecurity and cyber defense. These studies demonstrate LLMs' capabilities in strategic reasoning, IoT intrusion prediction, malware TTP analysis, enhancing BAS systems, synthesizing cybersecurity literature, and building Autonomous Cyber Defense agents. Collectively, they showcase LLMs' versatility in addressing various cybersecurity challenges, from threat detection to automated defense in diverse environments. However, while these contributions demonstrate significant advancements, a crucial gap remains in effectively leveraging LLMs to analyze comprehensive attack procedure databases and integrating this analysis into a SCE methodology. This unexplored approach could enable more sophisticated adversary behavior anticipation, allowing for the proactive design of adaptive security mechanisms that anticipate threats before they materialize, bridging critical gaps in the continuous development lifecycle.

\section{Large Language Model as an enabler of adversary anticipation}\label{method}

%Introduction
This section presents the proposed flow for generating attack-defense trees that represent potential adversary behavior, as well as the key metrics defined to evaluate the quality of the trees generated by LLMs. %The process flowchart is shown in Figure~\ref{fig:FlowDiagram}, and each step will described in the following sections.

\subsection{Generating an attack-defense tree}\label{sec:gen_attack_tree} 

\begin{figure*}[!th]
\centering
	\includegraphics[width=1\textwidth]{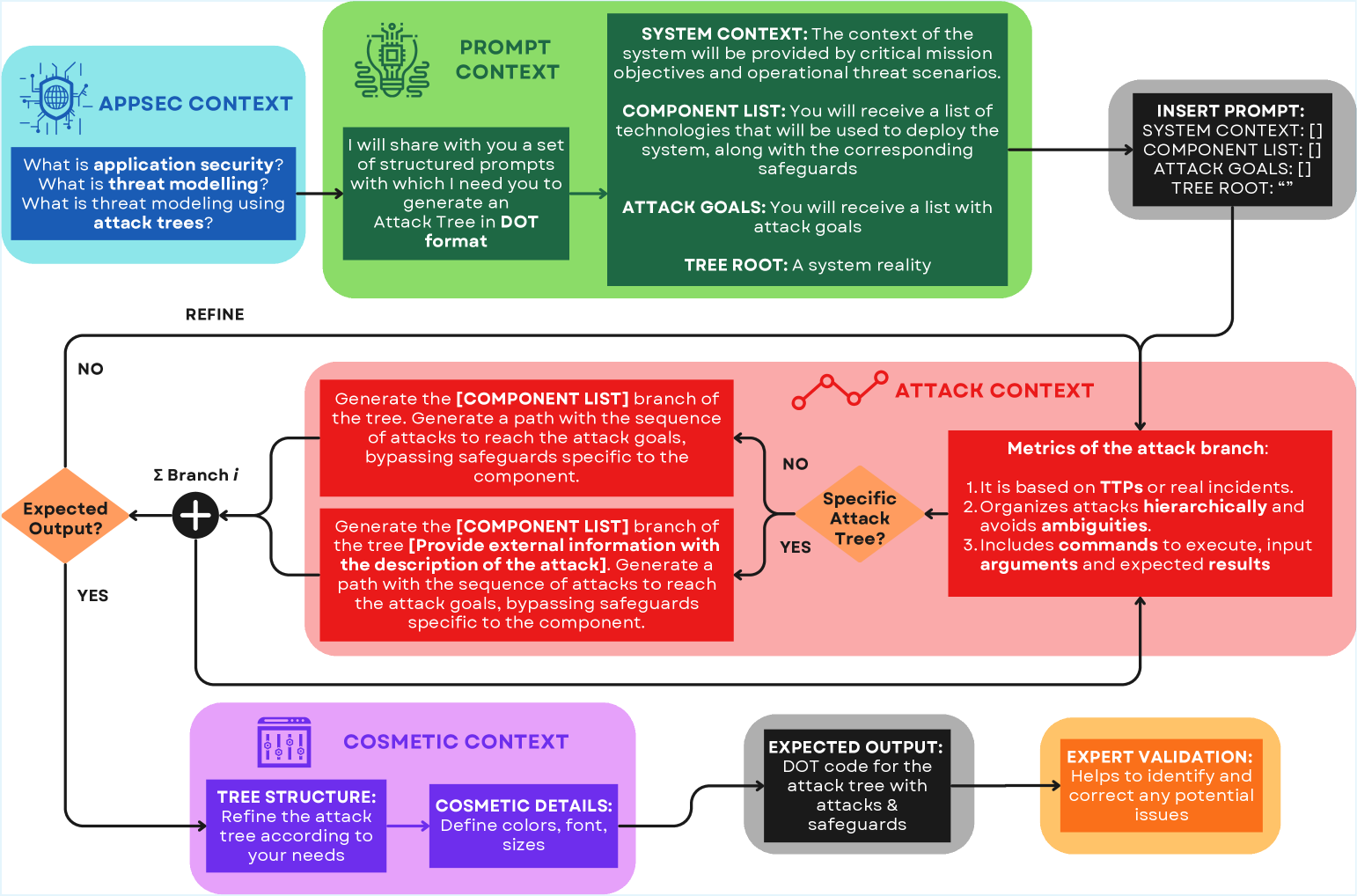} 
	\caption{Flow diagram of the construction of an attack-defense tree using LLM}
	\label{fig:FlowDiagram}
\end{figure*}

%Introduction to Attack-Defense tree 
Attack-defense trees provide a structured framework for modeling and analyzing adversarial interactions in cybersecurity scenarios by integrating both offensive and defensive actions into a single hierarchical representation. These trees extend traditional attack trees by incorporating defensive countermeasures, enabling analysts to visualize the dynamic interplay between attackers seeking to exploit vulnerabilities and defenders implementing mitigation. The visual nature of attack-defense trees also enhances interdisciplinary collaboration, enabling technical teams and stakeholders to jointly evaluate security postures and prioritize mitigation against multi-stage threats.

%Using attack-defense tree as a tool to anticipate adversary behavior in a cyber defense scenario
Consequently, in our proposal, the attack-defense trees are used as part of a cyber defense strategy, allowing an organization that uses such trees to evaluate its own critical vulnerabilities and predict adversary behavior patterns according to known attack procedures. As a result, such an organization can pose defense strategies that are adequate considering a cost-benefit analysis and the capacities of the adversary. For the generation of the attack-defense trees, we have provided the LLMs with structured, individual prompts focused on single tasks, rather than a single prompt containing multiple tasks, which offers significant advantages. 

LLMs tend to perform better when solving smaller, discrete tasks compared to handling multiple complex tasks simultaneously. This phenomenon can be attributed to the model's inference capabilities, which are constrained by the number of tokens it can process. Another benefit of using multiple individual prompts lies in the model's recall capability. This is particularly important as it enables the LLM to infer attacks and controls that are more contextually aligned with the specific system or infrastructure components being analyzed.

%Introduction to the proposed flow
Thus, we propose the flow diagram depicted in Figure~\ref{fig:FlowDiagram}, which represents the interactions that a cybersecurity analyst should have with an LLM to construct an attack-defense tree. Such a flow will be described next and is composed of the following 6 phases: \textit{Application Security Context, Prompt Context, Insert Prompt, Attack Context, Cosmetic Context, and Expert validation.}

\subsubsection{Phase 1: Application Security Context} %Mario
The flow depicted in Figure~\ref{fig:FlowDiagram} begins by establishing the context for the LLM model by introducing key concepts such as application security, threat modeling, and attack-defense trees. This step is crucial to ensure that the model remains focused on the relevant subject matter for future tasks and avoids generating content unrelated to these topics. To achieve this, it is possible to pose questions to the LLM about these areas and request summaries of the information. 

This approach helps consolidate concise and relevant knowledge within the model's memory, which is inherently limited. Questions like ``What is application security?'', ``What is threat modeling?'', and ``What is threat modeling using attack trees?'' serve as effective starting points.

\subsubsection{Phase 2: Prompt Context}%Mario
In the following phase, the LLM model is provided with structured prompts and generates the desired output in the form of an attack-defense tree using DOT format, which is a way to represent these trees as directed graphs using the Graphviz language \cite{graphviz}. The structured prompt format defined for this approach is as follows:
\begin{itemize}
    \item \textbf{System Context}: The context of the system is provided in a specific form, e.g., ``The context of the system will be provided by critical mission objectives and operational threat scenarios''.
    \item \textbf{Component List}: It contains a list of technologies that will be used to deploy the system, along with the corresponding safeguards for each technology, e.g., ``AWS EC2 (monitored by Amazon GuardDuty to detect anomalous accesses)''.
    \item \textbf{Attack Goals}: It refers to a phrase that indicates to the LLM that an attack goals is provided, e.g. ``Ex-filtrate supply chain information''.
    \item \textbf{Tree Root}: The main node of the tree is specified to the LLM, e.g., ``Cloud-based supply chain System''.
\end{itemize}

The decision to use a 4-parameter prompt was based on a series of tests, which revealed that including additional parameters led LLMs to produce meaningless code. It is important to note that the order of the parameters had no impact on the outcomes.

\subsubsection{Phase 3: Insert Prompt}
This phase is critical to the process as it requires summarizing all the knowledge about the system under development. The outcomes at this stage can vary depending on the LLM being used. 

\subsubsection{Phase 4: Attack Context}

%Metrics to be considered per branch
In order to achieve certain quality in the results produced by LLMs, it becomes necessary to establish a series of metrics. After reading the literature, we can determine that an attack branch should be realistic, meaning it should be based on documented TTPs (Tactics, Techniques, and Procedures) from frameworks such as MITRE ATT\&CK or real-world incidents. Additionally, the steps should be organized hierarchically (objectives, steps, and commands) and avoid ambiguities. Finally, the branches should include executable commands, adjustable parameters, e.g., IP addresses, resource names, among others, and expected outcomes.

%Generalized vs specific attack-defense tree
Thanks to the versatility of the LLM models, it is possible to generate both generalized and specific attack-defense branches. On one hand, generalized attack-defense branches offer a broad overview of possible attacks a system might face. However, it requires a security analyst's expertise to determine whether the inferred attacks are genuinely exploitable within the system. To create a generalized attack-defense branch, the security analyst must prompt the LLM to infer potential attacks based on the system's context, topological components, safeguards, and attack goals. 

On the other hand, a specific attack-defense branch provides a more detailed view of the exact sequence of attacks an adversary might follow to achieve their objectives. This approach is particularly useful for security analysts who need precise attack procedures. To create a specific branch, the security analyst must provide the LLM with resources containing a detailed attack, such as one of the attack procedures included in HackTricks~\cite{hacktricks}.

For better results, the security analyst can ask the LLM to generate branches independently for each system component and then combine them into a complete tree. Afterward, it is essential to verify whether the generated attacks align with the intended attack goals. If not, those branches should be refined by iterating the process and including the details needed inside the prompts.

\subsubsection{Phase 5: Cosmetic Context}
Once the security analyst is satisfied with the generated attack-defense tree, he/she can request the LLM to apply cosmetic adjustments, such as modifying the text font, color, size, or node colors, among other visual elements. Additionally, the analyst can instruct the LLM to redefine the tree's structure if necessary. This process may be required in cases where the generated attack-defense tree contains redundant connections between nodes or if the arrangement of parent nodes or attack objectives does not meet the analyst's expectations.

\subsubsection{Phase 6: Expert Validation}
Once the attack-defense tree is generated, it undergoes a thorough validation by a security analyst. If the output does not meet expectations, e.g., if it includes irrelevant attacks, fabricated controls, disconnected nodes, or lacks meaningful relationships between attacks and goals, the analyst can provide feedback. The tree can be returned to the LLM in DOT format with specific instructions for revisions. This expert validation process is crucial for identifying and addressing potential shortcomings, ensuring that the final output is accurate, reliable, and applicable to real-world scenarios.

\subsection{Measuring the quality of an attack-defense tree}\label{subsec:metrics} 

%Introduction to metrics 
In this proposal, these attack-defense trees are intended to support the construction of SCE experiments, and to achieve this objective, they must have a well-defined structure and avoid ambiguities that could delay or even prevent the subsequent validation of attack scenarios aimed at anticipating adversary behavior. To this end, the following metrics are proposed: Attacks based on known TTPs, Ordered attack procedures, and Usable attack procedures.

Attack-defense trees benefit from incorporating the first one as this grounds the model in real-world adversary behaviors, ensuring scenarios align with documented threat intelligence and reducing ambiguity during validation. Ordered attack procedures enforce logical sequences of attack steps, reflecting the dependencies and progression of real-world attacks, which is critical for structuring reproducible experiments and accurately anticipating adversarial workflows. Finally, usable attack procedures prioritize actionable, non-redundant steps that are feasible to implement in controlled experiments, avoiding overly complex or theoretical paths that could hinder practical validation. Together, these metrics enhance the clarity, realism, and operational relevance of attack-defense, enabling systematic testing of defenses against validated attack vectors while minimizing ambiguities in scenario execution.

\subsubsection{Attacks based on known TTPs} This metric ensures that the tree incorporates real TTPs used in established frameworks such as MITRE ATT\&CK or the Cyber Kill Chain framework, which are already well-structured. Thus, minimizing potential hallucinations by the models when generating attack paths.
\begin{equation}\label{eq:MITRE_Score}
MITRE_{Score} = \frac{\sum_{i=1}^{n} M_i}{n} \times 100  
\end{equation}
\begin{itemize}
    \item $M_i$ = Binary value (1 if node $i$ corresponds to a MITRE ATT\&CK technique appropriate to the attack being carried out, 0 otherwise)
    \item $n$ = Total number of attack nodes in the tree
\end{itemize}

\subsubsection{Ordered attack procedures} The branches of the tree must be organized hierarchically, starting from a root node that connects the attack nodes and ends in a target node. Additionally, attack paths must be clear and unambiguous, accurately reflecting the process an attacker follows to exploit a vulnerability:
\begin{equation}\label{eq:Ordered_Score}
Ordered_{Score} = (1-\frac{N_{d} + N_{sc}}{n}) \times 100    
\end{equation}

\begin{itemize}
    \item $N_{d}$ = Number of nodes deviated from the original order established in the external source attached to the LLM
    \item $N_{sc}$ = Number of nodes without children that are not a final node
    \item $n$ = Total number of attack nodes in the tree
\end{itemize}

\subsubsection{Usable attack procedures}
Includes commands to execute, input arguments and expected results. Attack and defense nodes must clearly describe what is done (procedure), how it is done (commands and input arguments), and the expected outcome:
\begin{equation}\label{eq:Usable_Score}
Usable_{Score} = \frac{\sum_{i=1}^n (C_i + I_i + R_i)}{3n} \times 100    
\end{equation}

\begin{itemize}
    \item $C_i$ = 1 if the node contains at least one real executable command appropriate to the attack being carried out, 0 otherwise.
    \item $I_i$ = 1 if the node contains input parameters necessary to execute the command, 0 otherwise.
    \item $R_i$ = 1 if the node contains descriptive expected results, 0 otherwise.
    \item $n$ = Total number of attack nodes in the tree.
\end{itemize}

Finally, we can compute the overall score of the tree with the following equation:
% \begin{equation}
% Tree\_Score = \frac{\ref{eq:MITRE_Score} + \ref{eq:Ordered_Score} + ~\ref{eq:Usable_Score}}{3}    
% \end{equation}
\begin{equation}
Tree_{Score} = \frac{MITRE_{Score} + Ordered_{Score} + Usable_{Score}}{3}    
\end{equation}

\section{Experiments}\label{experiments}

%Summary 
This section of experiments is composed in the following way: Section~\ref{subsec:IdsSettings} where the cyber defense scenario is established, Section~\ref{subsec:KDB}, where the attack-defense trees are generated, Section~\ref{subsec:SCEexperiment}, where one of the branches from the previously generated attack-defense trees is selected and translated into a SCE experiment, and Section~\ref{subsec:IdsAnalysis}, where the results obtained from the attack generation process are presented and analyzed.

\subsection{Settings}\label{subsec:IdsSettings}

%Definition of the scenario
This section demonstrates the practical implementation of the methodology proposed in Section~\ref{method} through a military sector case study motivated by the critical need to model and anticipate adversarial tactics in cyber defense planning. In this case study, a Ministry of Defense (MoD) operates a military logistic information system with a cloud-native approach, designed to get the most benefits of cloud capabilities while also being resilient enough to counter possible sophisticated threat actors. This military logistics system track essential war fighter supplies across global bases and handles classified data including ammunition, medical kits, and rations, which need to be protected against nation-state cyber espionage tactics. The military logistics system implements a security-first design philosophy which stems from historical analysis of adversaries behavior targeting military logistics systems.

%Details for AWS example
With the aim of defining a specific scenario for the experiments, it is assumed that the military logistics system is deployed in AWS GovCloud and employs three key managed services: EC2 for secure computing, CodeBuild for automated CI/CD pipelines, and CodeGuru for AI-powered code security analysis. This cloud architecture provides a realistic scenario for assessing the capabilities of LLMs in generating attack-defense trees based on the data flow presented in Section~\ref{method}. Building upon this foundation, two specific LLMs were selected for their evaluation in reasoning and cybersecurity applications: GPT-4 and QwQ-32B. These models will serve as the core components for simulating adversary behaviors and creating structured attack-defense models, further enhancing the experiment’s relevance and applicability to real-world cybersecurity scenarios. For the experiment, we have used the web versions of both models, available at \url{https://chat.qwen.ai} and \url{https://chatgpt.com}, respectively.

\subsection{Generating Attack-Defense trees} \label{subsec:KDB}

\begin{figure*}[!ht]
    \centering
    \includegraphics[width=1.7\columnwidth]{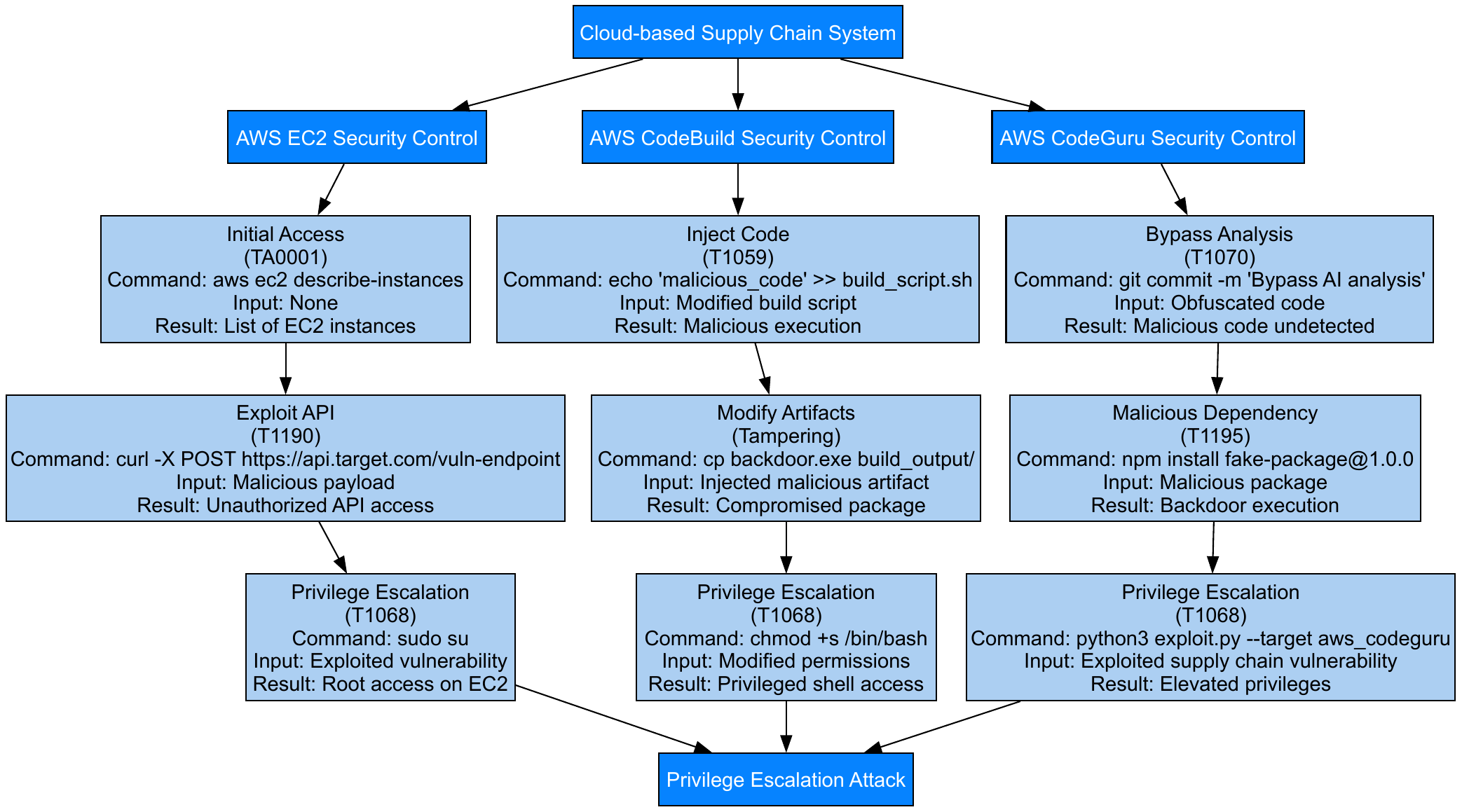}
    \caption{Reduced version of the attack-defense tree generated by GPT-4}
    \label{fig:attacktreeGPT}
\end{figure*}

\begin{figure*}[!ht]
    \centering
    \includegraphics[width=2\columnwidth]{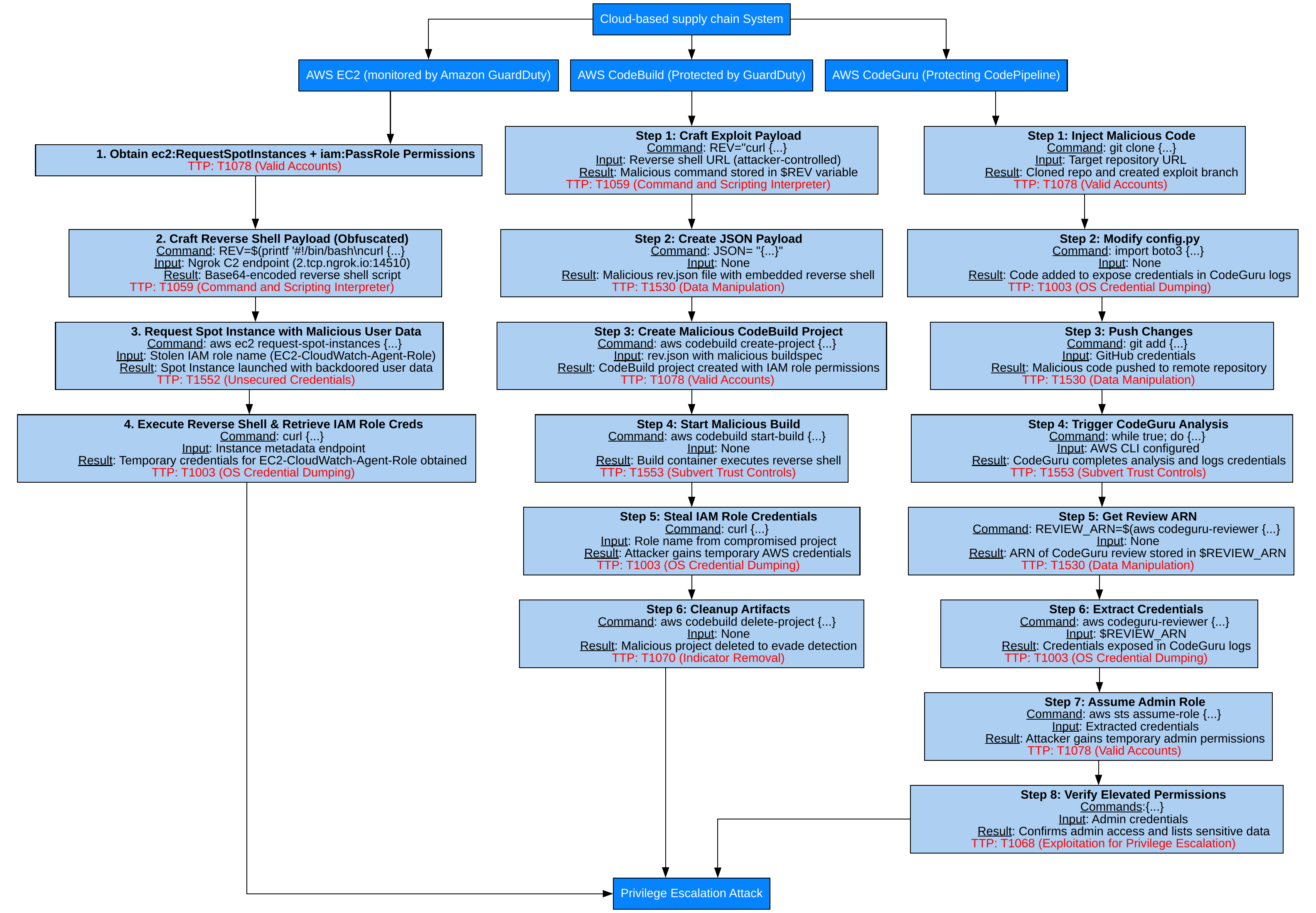}
    \caption{Reduced version of the attack-defense tree generated by QwQ-32B}
    \label{fig:attacktreeQwen}
\end{figure*}

The attack trees generated when applying the proposed iterative process with both models are shown in Figures ~\ref{fig:attacktreeGPT} and ~\ref{fig:attacktreeQwen}. Each branch begins with a node (represented in dark blue) that identifies the specific AWS service being targeted. From there, each branch follows a series of attack nodes (represented in light blue) that detail the specific steps, commands, and techniques used in the attack sequence. Despite the diversity in attack methodologies, all three paths ultimately converge toward the same goal: executing a privilege escalation attack.

\begin{itemize}
    \item \textbf{Branch 1:} In the first branch, i.e., EC2 attack branch, the attacker gains access to EC2 resources, executes a reverse shell to establish a connection with their systems, and steals sensitive credentials from the instance. This allows them to escalate privileges and access additional AWS resources.

    \item \textbf{Branch 2:} In the second branch, i.e., CodeBuild attack branch, the attacker uses valid credentials to set up a compromised CodeBuild project that runs a reverse shell, enabling them to steal credentials from the build environment. They then delete the project to cover their tracks while maintaining unauthorized access to AWS resources.

    \item \textbf{Branch 3:} In the third, i.e., CodeGuru attack branch, the attacker injects malicious code into a repository and manipulates configurations to expose sensitive information during automated code reviews. By extracting credentials from the analysis process, they escalate privileges and gain unauthorized access to sensitive data.
\end{itemize}

%Best Attack-Defense tree generated by GPT-4 
%Best Attack-Defense tree generated by Qwen %Mario

The trees generated by both LLMs were evaluated based on the three key metrics proposed in Section~\ref{subsec:metrics}. Overall, as we can appreciate in Table~\ref{tab:metricsComparison}, a comparison of final scores reveals that QWQ-32B achieved a $Tree_{Score}=71.60\%$, while GPT-4 scored a $Tree_{Score}=61.73\%$. In mapping real-world TTPs from the MITRE ATT\&CK framework, QwQ-32B shows a more reliable representation with minimal hallucination, compared to GPT-4, which often provides incomplete or incorrect information. Both models excel in creating hierarchical structures for attack procedures, but QwQ-32B offers clearer paths and priorities. Importantly, QwQ-32B outperforms GPT-4 in providing usable attack procedures, offering more detailed commands and expected outcomes. Having all this information considered, this makes QwQ-32B the preferred choice for subsequently detailing an SCE experiment from this tree.

\begin{table}[ht]
\resizebox{\columnwidth}{!}{%
\begin{tabular}{@{}lll@{}}
\toprule
Metric & QwQ-32B & GPT-4 \\ \midrule
\begin{tabular}[c]{@{}l@{}}Attacks Based on \\ Known TTPs\end{tabular} &
  \begin{tabular}[c]{@{}l@{}}Reliably maps real-world TTPs \\ from MITRE ATT\&CK \\ with minimal hallucination \\(22.22\%) (4/18 nodes).\end{tabular} &
  \begin{tabular}[c]{@{}l@{}}Most TTPs are \\ incomplete or wrong, \\ slightly reducing realism \\(11.11\%) (1/9 nodes).\end{tabular} \\ \midrule
\begin{tabular}[c]{@{}l@{}}Ordered Attack \\ Procedures\end{tabular} &
  \begin{tabular}[c]{@{}l@{}}Clear hierarchical structure \\ with unambiguous paths \\ and well-defined priority \\(100\%).\end{tabular} &
  \begin{tabular}[c]{@{}l@{}}Hierarchical but occasionally \\ introduces minor ambiguities \\ in node relationships \\(100\%).\end{tabular} \\ \midrule
\begin{tabular}[c]{@{}l@{}}Usable Attack \\ Procedures\end{tabular} &
  \begin{tabular}[c]{@{}l@{}}Highly detailed commands, \\ inputs, parameters, and \\ expected outcomes \\(92.59\%) (50/18 nodes).\end{tabular} &
  \begin{tabular}[c]{@{}l@{}}Provides usable steps \\ but lacks technical \\ depth in some areas \\(74.07\%) (20/9 nodes).\end{tabular} \\ \bottomrule
\end{tabular}%
}
\caption{Comparative table of obtained metrics}
\label{tab:metricsComparison}
\end{table}

\subsection{SCE experiment} \label{subsec:SCEexperiment}

%Intro to the experiment
The purpose of this experiment is to evaluate the resilience of an AWS environment to potential misuse of permissions. Specifically, the experiment will test whether an attacker can successfully request a Spot Instance, attach a privileged IAM role, and execute a reverse shell script to steal the IAM role credentials, using the permissions $ec2:RequestSpotInstances$ and $iam:PassRole$.

Based on the goal of the attack tree, i.e., perform a privilege escalation attack, it is possible to define the experiment following the SCE methodology:

\begin{itemize}
    \item \textbf{Observability}: Detection of Spot Instance creation with suspicious user data and anomalous permission usage via AWS GuardDuty.
    \item \textbf{Steady State}: A secure AWS environment where GuardDuty detects privilege escalation, misconfigured permissions are non-exploitable, and security controls function as designed.
    \item \textbf{Hypothesis}: AWS GuardDuty will detect privilege escalation attempts when Spot Instances with misconfigured permissions and malicious user data are launched.
\end{itemize}

\subsection{Experiment Methodology}\label{subsec:IdsAnalysis} 

This section describes the execution of the SCE Privilege Escalation experiment using a Spot EC2 instance. The automation of the experiment, which replicates the attack from branch \#1 of the attack tree shown in Figure~\ref{fig:attacktreeQwen}, culminates in obtaining temporary credentials for the $EC2-CloudWatch-Agent-Role$. Further detailed information about the experiment, along with the steps to replicate it, can be found in the following repository: \url{https://github.com/mariomc14/devsecops-adversary-llm.git}. Specific actions performed per each stage of the experiment are described next: 

\begin{itemize}
    \item GuardDuty Findings: The initial state of Amazon GuardDuty is verified, ensuring that no prior alerts exist before executing the subsequent phases of the experiment. If no alerts are detected, the execution proceeds with the creation of the Spot EC2 instance.
    \item Spot EC2 Instance Creation: The command to create the instance with the $EC2-CloudWatch-Agent-Role$ is executed, incorporating the Base64-encoded string as a payload in the UserData field. During the instance creation process, a connection for reverse shell listening is established in parallel. 
    %The corresponding command is shown in Figure [].
    \item Credential Extraction and Usage: In the third and final stage of the experiment, the obtained credentials are extracted and used by executing the request shown below. These credentials are stored and leveraged to perform actions that the attacker initially did not have permission for, such as listing users.
\end{itemize}

\section{Conclusions and Future Work}\label{conclusions}

%Summary of achievements
In this paper, we proposed an LLM-based flow for generating attack-defense trees that represent adversary behavior. We defined key metrics to evaluate the quality of the generated attack-defense trees. Based on the generated diagrams, we constructed an SCE experiment, relying on one of the previously selected branches of the attack-defense tree. Subsequently, we executed the SCE experiment in ChaosXploit, achieving privilege escalation in an EC2 Spot instance, as part of a military supply chain system in a cyber defense scenario. The results demonstrated the capability of the attack-defense trees generated by our proposed flow to anticipate adversary behavior in exploiting pre-existing system vulnerabilities, and its utility in a cyber defense strategy.

%Reduced future work
Future work will explore extending this framework to support automated countermeasure recommendations to enhance its practicality for anticipating adversary behavior in real-world scenarios.

\section*{Acknowledgment}
This work has been co-funded by the European Union (project ECYSAP EYE). Views and opinions expressed are however those of the author(s) only and do not necessarily reflect those of the European Union or the European Defence Fund. Neither the European Union nor the granting authority can be held responsible for them.
This work has also been partially supported by MCIN/AEI/10.13039/501100011033 NextGeneration EU/PRTR, UE, under Grant TED2021-129300B-I00, by MCIN/AEI/10.13039/501100011033/FEDER, UE, under Grant PID2021-122466OB-I00, by  the Spanish National Institute of Cybersecurity (INCIBE) by the Recovery, Transformation and Resilience Plan, Next Generation EU under the strategic project DEFENDER, by the CyberDataLab (Cybersecurity and Data Science Laboratory) at the University of Murcia (Spain), and the School of Engineering, Science and Technology at the University of Rosario (Colombia).

%%% inclusión de referncias
\bibliographystyle{IEEEtran}
\bibliography{bibliography.bib}{}

\end{document}